\documentstyle[prl,aps]{revtex}

\begin{document}
\title{Pion Form Factor and Ambiguities in a Renormalizable Version of the
Nambu-Jona-Lasinio Model\thanks{%
Supported by CNPq-Brazil}}
\author{B. M. Rodrigues, A. L. Mota\thanks{%
motaal@ufsj.edu.br},}
\address{{\normalsize Departamento de Ci\^{e}ncias Naturais, Universidade Federal de
S\~{a}o Jo\~{a}o del Rei}\\
{\normalsize Caixa Postal 110, CEP 36.300-000, S\~ao Jo\~ao del Rei, MG,
Brazil}}
\maketitle

\begin{abstract}
We analyze the presence of an ambiguity in the pion electromagnetic form
factor within a renormalizable version of the Nambu-Jona-Lasinio model. We
found out that the ambiguity present on the evaluation of the form factor
decouples from its transversal part, confirming previous results obtained
ignoring the ambiguity. This result helps us to understand the role played
by finite but undetermined quantities in Quantum Field Theories.
\end{abstract}

%
%

\section{Introduction}

Recently (\cite{Colladay},\cite{Kostelecky},\cite{Jackiw}) it was brougth to
attention the existence of finite but undetermined radiative corrections in
Quantum Field Theory. These finite amplitudes are related to differences
between divergent quantities of the same degree of divergence, and the
result obtained by employing a particular regularization scheme can be
different of the result obtained by another one (in particular, this
difference appears between gauge invariant and non invariant schemes)\cite
{Perez}. This situation becomes more complex in models with parity violating
quantities, where dimensional regularization is inappropriate. A classical
example is the ABJ anomaly\cite{ABJ}, where one must choose between the
transversality of the vector or axial-vector currents. At this point, the
ambiguity plays a crucial role in determining which one of the symmetry
relations will be violated.

Scarpelli et al. \cite{Scarpelli} studied this situation employing a
regularization independent procedure that have been called Implicit
Regularization (IR) (\cite{Orimar},\cite{Brizola}). By using IR, they had
shown that the same ambiguities can occur in several situations, as in QED
vacuum polarization tensor. In this case, gauge invariance has to be used to
fix the ambiguity value. As proposed in \cite{Jackiw}, the ambiguity was
fixed in order to preserve some physical relationship, e.g., symmetry. One
of the advantages of the IR procedure is that the amplitudes can be
evaluated almost all the time without specifying some particular
regularization scheme. Of course, particular attention must be paid to
regularization dependent quantities. Another procedure that have been
employed based on the same principles is the differential regularization 
\cite{DifferentialR}, where the amplitudes are evaluated in configuration
space. For our purposes here, i.e., evaluate the pion form factor in a
regularization independent way in energy-momentum space, IR is more
appropriate.

Non-renormalizable models do not show the presence of ambiguities, in the
sense that the regularization scheme employed is part of the model. This is
the case of the Nambu-Jona-Lasinio (NJL) model (\cite{Nambu},\cite{Klevansky}%
), and one can say that Pauli-Villars regularized NJL model is different of
the sharp covariant cut-off regularized NJL model. Once defined which
regularization scheme will be used on the amplitudes evaluation of the NJL
model, also the result of the differences between divergent quantities will
be defined. Thus, there are no reasons to leave the ambiguity undetermined
in such models. In general, gauge invariant regularization schemes are used
to treat the regularized NJL model, but regularization schemes that destroy
the quadratic divergence, necessary to do the correct adjustment of the
model to the phenomenology, must be avoided. This is the case of the usual
gauge invariant Pauli-Villars regularization, and some modifications of the
scheme must be done in order to preserve the quadratic divergence and
symmetries of the model\cite{Hiller}.

But in renormalizable extensions of the NJL model (\cite{NJLRen},\cite{Mota}%
) the regularization scheme is not part of the model, and thus these models
can present ambiguities. Due its relative simplicity, the renormalizable
extension of the scalar/pseudo-scalar section of the NJL SU(2) model
provides a good scenario where one can study the role played by ambiguities
in a QFT: (i) it is renormalizable; (ii) it is a fermionic model; (iii) it
presents chiral symmetry, in the vanishing fermion mass limit; (iv) it
presents parity violating couplings; (v) it presents ambiguities; (vi) there
are experimental data available to check its results. Of course there are
another models whose fulfill some of the characteristics listed above, such
as the chiral Schwinger model, the Gross-Neveu model, and so on.

We will adopt here the renormalizable extension of the NJL model presented
in \cite{Mota}. This renormalizable non-trivial extension is constructed by
using a mean field expansion \cite{Eguchi} that results in a renormalizable
but trivial effective Lagrangean \cite{Guralnik}. The first-order effective
Lagrangean presents kinetic and mesonic interaction terms radiatively
generated, and by augmenting the original effective lagrangean with similar
terms, one can avoid the non-renormalizability and triviality of the model 
\cite{Mota}. This procedure, of course, leads to a model that is different
from the original regularized NJL model, but that is related to the latter.
In this condition, results are formally identical to the results presented
by the NJL model, but without any explicit cut-off dependence (the
connection limit can be taken), leaving the model regularization scheme's
independent.

Both regularized and renormalized versions of the NJL model are appropriated
to the study of mesonic properties, and reproduces observables with a
reasonable agreement with experimental data. As discussed before, there is
not ambiguities in regularized NJL model, since they are fixed by the choice
of an specific regularization scheme. But in the renormalizable extension of
the NJL model, these ambiguities are present on the calculation of several
processes. The results presented in \cite{Mota} do not took in account the
influence of these ambiguities, they were fixed by implicitly employing a
gauge invariant regularization scheme that fixes them to zero. In this
letter, in particular, we will evaluate the pion electromagnetic form factor
within the renormalizable extension of the NJL model employing IR and,
following the prescription suggested in \cite{Jackiw}, leaving the ambiguity
undetermined. We will show that the ambiguity present on the pion form
factor cannot be fixed by symmetry relationships (Ward Identity) and does
not affect the previsibility of the model.

\section{The Pion Charge Form Factor}

In the scalar/pseudo-scalar sector of the NJL model, the pion form factor is
obtained by evaluating the diagram depicted in figure 1. There is a whole
contribution of the vector/axial-vector sector that reproduces the influence
of the vector mesons (Vector Dominance Model - VDM) that is missing in the
present analysis. In the time-like region the contribution coming from these
vector mesons is very important, so we will restrict our analysis to the
space-like region. We must also remark that our interest is to study the
presence of ambiguities in this process, and thus our results will not be
jeopardized by not employing the VDM.

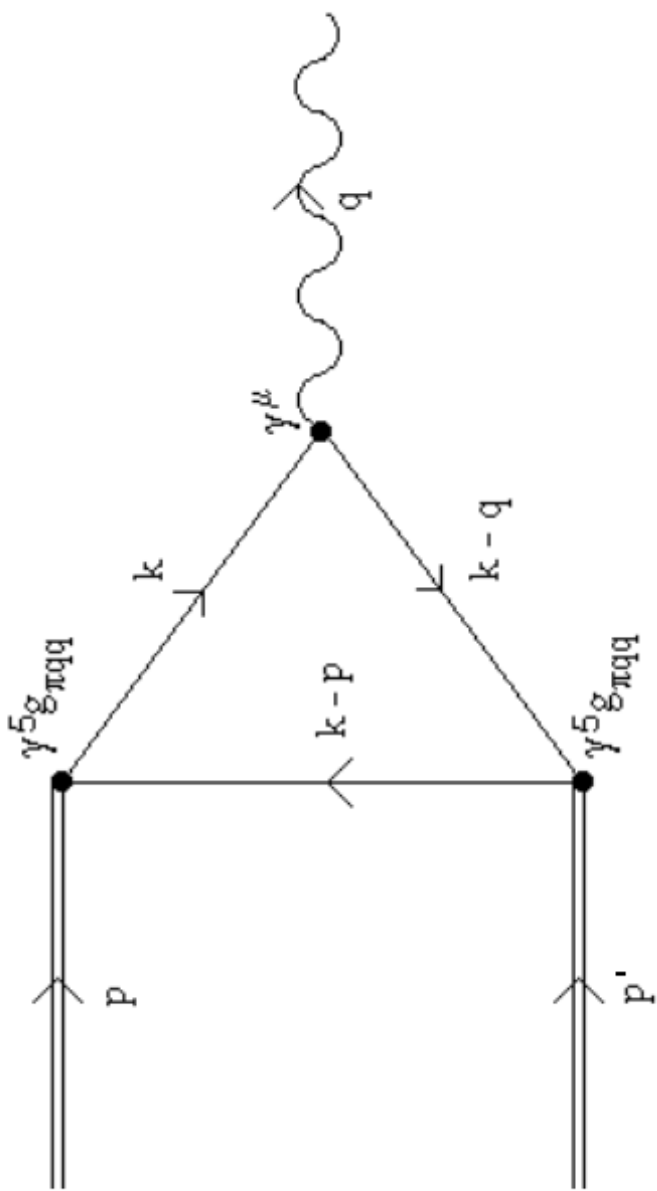
\begin{figure}[h]
\caption{Feynman diagram for the pion charge form factor in the NJL model.}
\label{diagram}
\end{figure}

The amplitude correspondent to the diagram showed in figure \ref{diagram} is
given by 
\begin{equation}
-i\Gamma ^\mu (p,p^{\prime })=N_cN_f\,g_{\pi qq}^2\int \frac{d^4k}{(2\pi )^4}%
Tr\{\gamma ^5\frac i{(\not{k}-\not{p})-m}\gamma ^5\frac i{\not{k}-m}\gamma
^\mu \frac i{(\not{k}-\not{q})-m}\}.  \label{Amplitude1}
\end{equation}
with $q=p+p^{\prime }$.

Evaluating (\ref{Amplitude1}) by using IR corresponds to isolate, without
the use of any explicit regularization, the divergences of the amplitude in
terms that are independent of the external momentum. As remarked in \cite
{Scarpelli}, we must not make use of symmetrical integration. By proceeding
in this way, we obtain (see notes on Appendix) 
\begin{eqnarray}
\Gamma ^\mu &=&2N_Cg_{\pi qq}^2(p^\mu -p^{\prime \mu })\frac 1{4N_Cg_\pi ^2}%
+2N_Cg_{\pi qq}^2m_\pi ^2(p^\mu -p^{\prime \mu })\left( Z_0^{\prime }(m_\pi
^2)-I_3(q^2,m_\pi ^2,m_\pi ^2)\right)  \label{Amplitude2} \\
&&-\,2N_Cg_{\pi qq}^2q_\nu Y^{\mu \nu },  \nonumber
\end{eqnarray}

where 
\begin{eqnarray}
Z_0(p^2) &=&\frac 1{16\pi ^2}\int_0^1dz\ell n\left[ 1-\frac{p^2}{m^2}%
z(1-z)\right] \\
&=&-\frac{2i}{(4\pi )^2}\{-1+\frac{\sqrt{p^4-4p^2m^2}}{p^2}\arctan h(\frac{%
p^2}{\sqrt{p^4-4p^2m^2}})\}  \nonumber
\end{eqnarray}

\begin{equation}
Z_0^{\prime }(p^2)=\frac{dZ_0(p^2)}{dp^2},
\end{equation}
and 
\begin{equation}
I_3(p,q)=i\int \frac{d^4k}{(2\pi )^4}\frac 1{%
(k^2-m^2)[(k-p)^2-m^2][(k-q)^2-m^2]}{.}
\end{equation}

At this point, we identify the ambiguity in (\ref{Amplitude2}) by the term 
\begin{equation}
Y^{\mu \nu }=2\int \frac{d^4k}{(2\pi )^4}\frac{k^\mu k^\nu }{(k^2-m^2)^3}%
-g^{\mu \nu }\int \frac{d^4k}{(2\pi )^4}\frac 1{(k^2-m^2)}.
\label{Ambiguity}
\end{equation}
It is a simple matter to evaluate the quantity in (\ref{Ambiguity}) in
different regularization schemes. So, in dimensional regularization or gauge
invariant Pauli-Villars one can obtain $Y^{\mu v}=0$, but in covariant sharp
cut-off one can obtain $Y^{\mu \nu }=\frac{ig^{\mu \nu }}{2(4\pi )^2}$.
Since we are not specifying here any explicit regularization scheme, we will
adopt the procedure suggested in \cite{Jackiw} and leave this ambiguity
undetermined up to the end of the evaluation. Setting, by using Lorentz
invariance, 
\begin{equation}
Y^{\mu \nu }=g^{\mu \nu }\alpha
\end{equation}
we can write (\ref{Amplitude2}) as 
\begin{eqnarray}
\Gamma ^\mu &=&2N_Cg_{\pi qq}^2(p^\mu -p^{\prime \mu })\frac 1{4N_Cg_\pi ^2}
\label{Amplitude3} \\
&&+2N_Cg_{\pi qq}^2m_\pi ^2(p^\mu -p^{\prime \mu })\left( Z_0^{\prime
}(m_\pi ^2)-I_3(p^2,m_\pi ^2,m_\pi ^2)\right)  \nonumber \\
&&-\,2N_Cg_{\pi qq}^2q^\mu \alpha .  \nonumber
\end{eqnarray}

The divergent terms present in (\ref{Amplitude3}) were absorbed in the
contra-terms of the model, since it is renormalizable, in the same way
presented in \cite{Mota}, resulting in renormalized parameters $g_\pi $ and $%
g_{\pi qq}$.

\section{The Ambiguity}

Equation (\ref{Amplitude3}) shows us that the ambiguity belongs to the
longitudinal term of the amplitude $\Gamma ^\mu $, the transverse part being
free of ambiguities. In order to verify if symmetries (such as gauge
invariance) fix the ambiguity, we must verify the Ward identity related to
the pion-photon vertex function, that in the NJL model reads 
\begin{equation}
q_\mu \Gamma ^\mu =g_{\pi qq}^2(T^{\pi \pi }(p,q)-T^{\pi \pi }(p))
\label{Ward}
\end{equation}
where 
\[
T^{\pi \pi }(p,q)=\int {\frac{d^4k}{(2\pi )^4}}Tr\{\gamma _5\frac 1{\not{k}-%
\not{p}-m}\gamma _5\frac 1{\not{k}-\not{q}-m}\} 
\]
and 
\[
T^{\pi \pi }(p)=\int {\frac{d^4k}{(2\pi )^4}}Tr\{\gamma _5\frac 1{\not{k}-m}%
\gamma _5\frac 1{\not{k}-\not{p}-m}\}. 
\]

In the left hand side of eq. (\ref{Ward}), the product of the photon
momentum $q_\mu $ by the transverse part of $\Gamma ^\mu $ vanishes. Thus,
only the ambiguous part of $\Gamma ^\mu $ survives, leading to 
\begin{equation}
2N_Cg_{\pi qq}^2q^2\alpha =g_{\pi qq}^2(T^{\pi \pi }(p,q)-T^{\pi \pi }(p))
\label{Ward2}
\end{equation}

The right hand side of eq. (\ref{Ward2}) can be explicitly evaluated, and,
also applying IR, the Ward identity (\ref{Ward}) can be exactly verified.
Thus, the Ward identity (\ref{Ward}) is satisfied even in the presence of
the ambiguity (\ref{Ambiguity}), and does not fix its value. This ensures
charge conservation as a feature independent of the presence of the
ambiguity.

The next question to be answered is how the ambiguity present in the pion
electromagnetic vertex function $\Gamma ^\mu $ affects the pion form factor.
Let us recall that the amplitude corresponding to figure \ref{diagram} can
be obtained by 
\begin{equation}
M=\lim_{p^2,p^{\prime 2}=m_\pi ^2}\varepsilon _{1\mu }\Gamma ^\mu
(p,p^{\prime })
\end{equation}
where $\varepsilon _{1\mu }$ is the polarization vector of the off-shell
photon with momentum $q=p+p^{\prime }$. Since $\varepsilon _{1\mu }$ is
transverse to the photon momentum, the longitudinal part of $\Gamma ^\mu $
will not be present in the amplitude. Thus, this amplitude and, in
consequence, the pion charge form factor, will be free from the ambiguity
present in $\Gamma ^\mu (p,p^{\prime })$.

\section{Numerical Results}

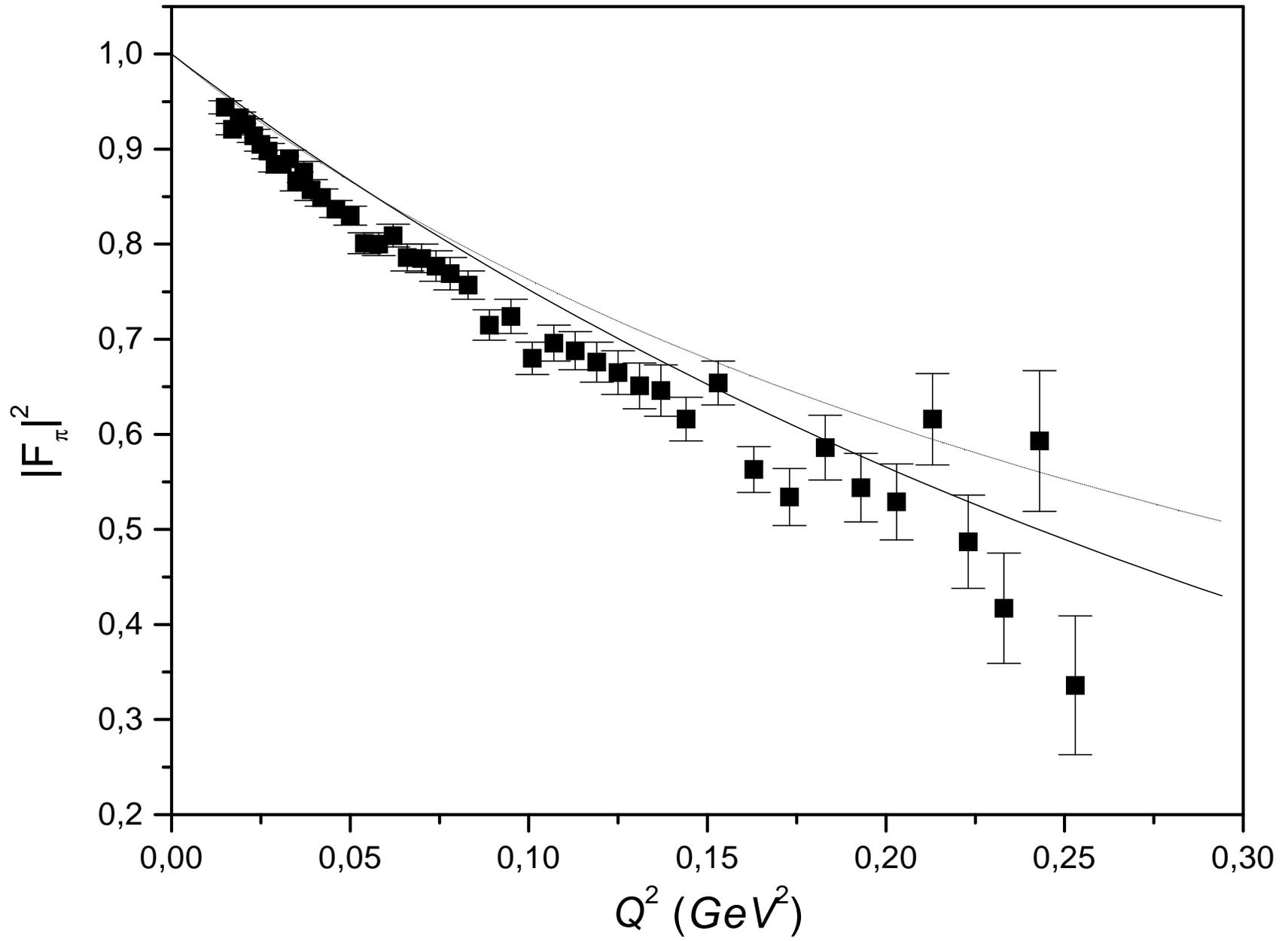
\begin{figure}[h]
\caption{Pion charge form factor in the space--like region. The results
obtained from the renormalizable extension of the NJL model are presented
for $m=210MeV$ (dotted line) and $m=350MeV$ (solid line) and are compared to
experimental data \protect\cite{Amendolia}. }
\label{chargeff}
\end{figure}

Although the ambiguity does not affect the pion form factor, we present here
the numerical results obtained for the pion form factor within the
renormalizable extension of the NJL model. The results and the parameters
set presented here are the same as in \cite{Mota}, we reproduce these
results for reasons of completeness. We apply implicit regularization, and
fit the following sets of parameters: ($m=350MeV,\,\,g_\pi =3.752$ and $\mu
_\pi =141MeV$) and ($m=210MeV,\,\,g_\pi =2.25$ and $\mu _\pi =141MeV$). The
parameters are adjusted to reproduce $f_\pi =93.3MeV$ and $m_\pi =139MeV$.
The comparison between the results of the renormalizable extension of the
NJL model and experimental data are shown in figure \ref{chargeff}. The
electromagnetic radius of the pion can be obtained by 
\begin{equation}
<r_\pi ^2>=-6\frac{dF_\pi (q^2)}{dq^2}|_{q^2=0}
\end{equation}
and, for the sets of parameters above we obtain, respectively 
\begin{equation}
<r_\pi ^2>=0.58fm;\,\,\,<r_\pi ^2>=0.6fm
\end{equation}
that are to be compared with the experimental result \cite{Amendolia} 
\begin{equation}
<r_\pi ^2>_{\exp }=0.678\pm 0.012fm
\end{equation}
We observe that a better fit is obtained with the lower constituent quark
mass, $m=210MeV$. In fact, if the constituent quark mass is lowered, the fit
with the experimental data is improved. But, as stated before, the model
studied here does not include the vector/axial-vector meson sector, and a
complete agreement with experiment is not to be expected.

\section{Conclusion}

In summary, we obtained the pion charge form factor within a renormalizable
extension of the SU(2) NJL model with scalar/pseudo-scalar couplings. We had
shown that, in this model, the pion-photon vertex function contains an
ambiguous term, i.e., a finite but regularization dependent term. We verify
that, even in the presence of the ambiguity, the Ward identity related to
this vertex function is satisfied, so it does not fix the ambiguity value.
Also, in the amplitude related to this process, and consequently in the pion
charge form factor, the ambiguity is not present, since it is transverse to
the photon polarization vector, and thus decouples from the physical content
of the model.

This provides an example of one of the roles played by ambiguities in
Quantum Field Theory: in previous works, it was shown that ambiguities can
be fixed either by symmetry relationships (as in QED\cite{Scarpelli}) or by
phenomenology (as in the neutral pion electromagnetic decay or in proton
decay\cite{Jackiw}). Here, we found out that there is a third situation,
where the ambiguity cannot be fixed by symmetry, but completely decouples
from the physical content of the model, remaining independent from the
regularization scheme employed.

\section{Acknowledgments}

This work was supported by CNPq-Brazil.

\section{Appendix}

In order to show explicitly how we isolated the ambiguity in the pion charge
form factor, we will proceed, in this appendix, the computation of the
ambiguous part of (\ref{Amplitude1}). After taking the traces on color,
flavor and Dirac spaces, we obtain the following two ambiguous integrals 
\begin{equation}
\xi ^{2\mu }=\int \frac{d^4k}{(2\pi )^4}\frac{k^2k^\mu }{%
(k^2-m^2)[(k-p)^2-m^2][(k-q)^2-m^2]},  \eqnum{A1}
\end{equation}
and 
\begin{equation}
p_\mu \xi ^{\mu \nu }=p_\mu \int \frac{d^4k}{(2\pi )^4}\frac{k^\mu k^\nu }{%
(k^2-m^2)[(k-p)^2-m^2][(k-q)^2-m^2]}.  \eqnum{A2}
\end{equation}

By adding and subtracting a $m^2k^\mu $ term on the numerator of (A1), we
obtain 
\begin{eqnarray}
\xi ^{2\mu } &=&\int \frac{d^4k}{(2\pi )^4}\frac{k^\mu }{%
[(k-p)^2-m^2][(k-q)^2-m^2]}  \eqnum{A3} \\
&&+m^2\int \frac{d^4k}{(2\pi )^4}\frac{k^\mu }{%
(k^2-m^2)[(k-p)^2-m^2][(k-q)^2-m^2]}  \nonumber \\
&=&I_t^\mu +m^2\xi ^\mu   \nonumber
\end{eqnarray}

The last integral in (A3) is finite, and can be evaluated directly, yielding 
\begin{equation}
\xi ^\mu =(p^\mu -p^{\prime \mu })(Z_0^{\prime }(m_\pi ^2)-I_3(q^2,m_\pi
^2,m_\pi ^2))  \eqnum{A4}
\end{equation}

Introducing one Feynman parameter on $I_t^\mu $, we obtain 
\begin{equation}
I_t^\mu =\int_0^1dx\int \frac{d^4k}{(2\pi )^4}\frac{k^\mu }{[(k-t)^2-M^2]^2}
\eqnum{A5}
\end{equation}

where 
\begin{equation}
M^2=m^2-(p-q)^2x(1-x),  \eqnum{A6}
\end{equation}
and 
\begin{equation}
t^\mu =p^\mu x+q^\mu (1-x).  \eqnum{A7}
\end{equation}

By adding and subtracting a $t^\mu $ term in the numerator of (A5), we
obtain 
\begin{equation}
I_t^\mu =\int_0^1dx\left\{ \int \frac{d^4k}{(2\pi )^4}\frac{k^\mu -t^\mu }{%
[(k-t)^2-M^2]^2}+t^\mu \int \frac{d^4k}{(2\pi )^4}\frac 1{[(k-t)^2-M^2]^2}%
\right\}   \eqnum{A8}
\end{equation}

The last integral in (A8) is logarithmically divergent, an we can safely
proceed a shift in the integration momentum $k-t\rightarrow k^{\prime }$.
More careful must be taken in the evaluation of the first integral in (A8),
since it is linearly divergent. To properly isolate the ambiguity term
present in this integral, we introduce the translation operator $e^{-t^\mu 
\frac \partial {\partial k^\mu }}$ and rewrite (A8) as 
\begin{equation}
I_t^\mu =\int_0^1dx\left\{ \int \frac{d^4k}{(2\pi )^4}e^{-t^\mu \frac 
\partial {\partial k^\mu }}\left\{ \frac{k^\mu }{(k^2-M^2)^2}\right\} +t^\mu
\int \frac{d^4k}{(2\pi )^4}\frac 1{[k^2-M^2]^2}\right\} .  \eqnum{A9}
\end{equation}

Expanding the translation operator, we find 
\begin{eqnarray}
\int \frac{d^4k}{(2\pi )^4}e^{-t^\mu \frac \partial {\partial k^\mu }%
}\left\{ \frac{k^\mu }{(k^2-M^2)^2}\right\} &=&\int \frac{d^4k}{(2\pi )^4}%
\frac{k^\mu }{(k^2-M^2)^2}-t^\nu \int \frac{d^4k}{(2\pi )^4}\frac \partial {%
\partial k^\nu }\left\{ \frac{k^\mu }{(k^2-M^2)^2}\right\}  \eqnum{A10} \\
&&+\frac{t^\nu t^\rho }2\int \frac{d^4k}{(2\pi )^4}\frac{\partial ^2}{%
\partial k^\nu \partial k^\rho }\left\{ \frac{k^\mu }{(k^2-M^2)^2}\right\}
+...  \nonumber
\end{eqnarray}

The integrals with odd terms in its integrands in (A10) vanish, and all the
terms with derivatives of order greater than 2 will also vanish. We obtain 
\begin{equation}
I_t^\mu =\int_0^1dx\left\{ \int \frac{d^4k}{(2\pi )^4}\frac{-2g^{\mu \nu }}{%
(k^2-M^2)^2}+\int \frac{d^4k}{(2\pi )^4}\frac{8k^\mu k^\nu }{(k^2-M^2)^3}%
\right\} +t^\mu \int \frac{d^4k}{(2\pi )^4}\frac 1{[k^2-M^2]^2}.  \eqnum{A11}
\end{equation}

We identify the terms in brackets in (A11) as the ambiguous part of $\xi
^{2\mu }$. Nevertheless, there is still a dependence on the integrand of
(A11) on $p$, $q$ and $x$, and one can argue if there is some finite
contribution coming from (A11). We remark that this term is ambiguous, and
the sum of an ambiguity with any finite non-ambiguous term will remain
ambiguous. Although this fact, one can find, after successively applying the
following relationship 
\begin{equation}
\frac 1{k^2-M^2}=\frac 1{k^2-m^2}+\frac{M^2-m^2}{(k^2-m^2)(k^2-M^2)} 
\eqnum{A12}
\end{equation}
and some straightforward calculations, that 
\begin{eqnarray}
&&\int_0^1dx\left\{ \int \frac{d^4k}{(2\pi )^4}\frac{-2g^{\mu \nu }}{%
(k^2-M^2)^2}+\int \frac{d^4k}{(2\pi )^4}\frac{8k^\mu k^\nu }{(k^2-M^2)^3}%
\right\}   \eqnum{A13} \\
&=&\int \frac{d^4k}{(2\pi )^4}\frac{-2g^{\mu \nu }}{(k^2-m^2)^2}+\int \frac{%
d^4k}{(2\pi )^4}\frac{8k^\mu k^\nu }{(k^2-m^2)^3}=4Y^{\mu \nu }.  \nonumber
\end{eqnarray}

The evaluation of $\xi ^{\mu \nu }$ follows almost the same steps we follow
in the evaluation of $\xi ^{2\mu }$. Since it is contracted with the
momentum $p$, we made use of 
\begin{equation}
k.p=\frac 12[(k^2-m^2)+p^2+((k-p)^2-m^2)]  \eqnum{A10}
\end{equation}
and obtain 
\begin{equation}
p_\mu \xi ^{\mu \nu }=\frac 12\{p^2\xi ^\mu +I_t^\mu -I_q^\mu \}, 
\eqnum{A11}
\end{equation}
where $\xi ^\mu $ is given by (A4), $I_t^\mu $ is given by (A13) and 
\begin{equation}
I_q^\mu =\int \frac{d^4k}{(2\pi )^4}\frac{k^\mu }{(k^2-m^2)[(k-q)^2-m^2]}, 
\eqnum{A12}
\end{equation}
can be computed in the same way we proceeded on the computation of $I_t^\mu $%
.

%
%

\end{document}